\documentclass[prd,floatfix,preprintnumbers,showpacs,superscriptaddress,nofootinbib]{revtex4}

\usepackage{epsf,epsfig}
\usepackage{graphicx}

\begin{document}

\title{Constraining Cosmic Topology with CMB Polarization}

\author{Alain Riazuelo}
\email{riazuelo@iap.fr}
\affiliation{Institut d'Astrophysique de Paris, 98bis boulevard 
Arago, F--75014 Paris, France}

\author{Samuel Caillerie}
\email{cailleri@discovery.saclay.cea.fr}
\affiliation{CE-Saclay, DSM/DAPNIA/Service d'Astrophysique, F--91191 
Gif-sur-Yvette Cedex, France}

\author{Marc Lachi\`eze-Rey}
\email{marclr@discovery.saclay.cea.fr}
\affiliation{CE-Saclay, DSM/DAPNIA/Service d'Astrophysique, F--91191 
Gif-sur-Yvette Cedex, France}

\author{Roland Lehoucq}
\email{roller@discovery.saclay.cea.fr}
\affiliation{CE-Saclay, DSM/DAPNIA/Service d'Astrophysique, F--91191 
Gif-sur-Yvette Cedex, France}

\author{Jean-Pierre Luminet}
\email{jean-pierre.luminet@obspm.fr}
\affiliation{Laboratoire Univers et Th\'eories, CNRS-UMR 8102, 
Observatoire de Paris, F--92195 Meudon Cedex, France}

\begin{abstract} 
Multiply connected space sections of the universe on a scale smaller
than the horizon size can leave an imprint on cosmic microwave
background polarization maps, in such a way that the so-called
``circles-in-the-sky'' method can be used to detect or constrain the
topology. We investigate some specific cases, namely toroidal and 
sixth-turn spaces, in order to show the influence of topology on CMB 
polarization. The correlation between matched points happens to be 
always positive and higher than 75\% regardless of the angular scale 
and of the cosmological parameters, except for reionization. This
figure is better than what occurs in temperature maps, but is achieved
only in the absence of noise.  It is only slightly reduced by the
filtering scheme.
\end{abstract}

\maketitle

\section{Introduction}

The question of a possible multiply connected topology of our cosmic
space goes back to the pioneering works of Schwarzschild and Friedmann
more than 75 years ago. In the past two decades, various strategies
and methods have been devised to probe a non trivial topology of the
space sections of the universe, using current or forthcoming data from
cosmological observations~\cite{revmeth}. Since the topological length
scales are a priori not known, it is useful to survey the largest
observable scale in order to increase the odds of detecting, or at
least constraining, the space topology. The Cosmic Microwave
Background (CMB) emitting region, the so-called last scattering
surface (LSS) is situated at a redshift of $z \simeq 1100$ and
represents the deepest region that can be studied in the
electromagnetic domain. Thus its usefulness for cosmology in general
is tremendous~\cite{revcmb}. Since more than one decade~\cite{cobe},
temperature anisotropies in the CMB radiation have been mapped as a
function of the direction. They result from density fluctuations in
the primordial plasma, which are the seeds of evolved structures in
the universe. Their study provides crucial information on the
matter-energy contents of the universe, as well as some hints on the
physical process which could have generated the first density
perturbations at a much earlier epoch~\cite{revcmb}.

CMB anisotropies can also be used to constrain the topology.  The main
imprint of a non trivial topology on the CMB is well-known in the case
when the characteristic topological length scale (called the
injectivity radius) is smaller than the radius of the last scattering
surface: the crossings of the LSS with its topological images give
rise to pairs of matched circles of equal radii, centered at different
points on the CMB sky, and exhibiting correlated patterns of
temperature variations~\cite{topocmb}. Such ``circles-in-the-sky''
searches are currently in progress using the WMAP data. They are
however computationally very expensive, and present results are not
completely clear. On the one hand, a massive search for matching
circles with radii larger than $25^\circ$ gave negative
results~\cite{circglenn}. But this result may well be overstated
since, on the other hand, two other more specific searches gave hints
for a positive detection. Roukema et al.~\cite{circrouk} extended the
search to smaller radii and claimed to have found six pairs of
antipodal matched circles in a dodecahedral pattern; Aurich et al.
\cite{aurich2005} also found a marginal hint for spherical spaces,
both searches being consistent with the Poincar\'e dodecahedron space
model recently proposed by some of us~\cite{lum03} to account for the
observed anomalies of the CMB angular power spectrum on large
scales. The statistical significance of such results still has to be
clarified. In any case, a lack of nearly matched circles does not
exclude a multiply connected topology on scale less than the horizon
radius: detectable topologies may produce circles of small radii which
are statistically hard to detect and current analysis of CMB sky maps
could have missed even antipodal matching circles because various
effects may damage or even destroy the temperature
matching~\cite{flatspaces,aurich04}. Moreover, even if it might
already seem severely constrained by observational data, there are
still unexpected and still unexplained features on the large angular
scales of the CMB temperature map~\cite{anomisocmb} which may hint for
some sort of breaking in statistical isotropy of the CMB, a feature
which is not shared by many models other than multiply connected
topologies, and which is present even when the injectivity radius is
larger than the size of the observable universe~\cite{flatspaces}.

Recent work by Aurich {\it et al.}~\cite{aurich04},
Gundermann~\cite{gunder05} and Caillerie {\it et
al.}~\cite{caillerie05} support the dodecahedron model. Since the
above mentioned anomalies detected in the first year WMAP
data~\cite{anomisocmb} suggest the presence of some statistical
anisotropy in the CMB radiation, it appears thus necessary to better
constrain topology, by using tests of different nature. It is now
widely understood that the polarization of the CMB, predicted long
ago~\cite{Rees68}, can provide a lot of additional informations for
reconstructing the cosmological model~\cite{HuWhite97}. The
information encoded in polarization is expected to become a necessary
input for precision cosmology, when it will be mapped in detail in the
next release of the WMAP data~\cite{wmaphttp} and later by the Planck
satellite mission~\cite{planckhttp}. In this article we show how the
polarization can also be used to put additional constraints on space
topology.

This paper is organized as follows: in section~\ref{Seccorr}, we
remind the key ingredients for detecting topological signatures in a
CMB temperature map, using the circles-in-the-sky method. In
section~\ref{Secqr}, we recall the basic properties of CMB
polarization that are useful for our purpose.  We then compute in
section~\ref{SecPol} the expected correlation for polarization maps,
taking into account the specificities of polarization. Our main result
is that the correlation in polarization maps is always larger than
75\%, assuming an ideal case where no noise is present in the maps. In
section~\ref{SecSim} we present simulated maps for the 3-torus and the
sixth-turn space in order to illustrate the validity of our method. In
section~\ref{SecBlur}, we look for various sources of blurring of the
correlation when considering two effects: reionization and finite
resolution.

\section{CMB physics and correlation between matched points in 
temperature maps}
\label{Seccorr}

Let us first review the different contributions to CMB temperature as
well as polarization anisotropies. The apparent temperature
fluctuation in a given direction $\hat {\bf n}$ can be expressed (in
Newtonian gauge) by
\begin{equation}
\Theta(\hat {\bf n}) \equiv \frac{\delta T}{T}(\hat {\bf n}) =
   (\left. \frac{1}{4}\frac{\delta \rho}{\rho} + \Phi) \right|_{r 
\hat {\bf n}}
 - \hat {\bf n} . {\bf v}_e(r \hat{\bf n})
 + \int_0^{r} \left.(\dot \Phi + \dot \Psi)\right|_{l \hat{\bf n}}
              {\rm d} l ,
\label{TempFluc}
\end{equation}
where the quantities $\Phi$ and $\Psi$ are the usual Bardeen
potentials~\cite{bardeen}, ${\bf v}_e$ being the velocity within the
electron fluid. The first terms are evaluated at the LSS, i.e., at $r
\hat{\bf n}$, where $r$ represents the radius of the LSS. They
represent the Sachs-Wolfe and Doppler contributions to CMB temperature
anisotropies. The last term give accounts of the energy exchanged by
photons with time-varying gravitational fields, known as the
integrated Sachs-Wolfe (ISW) effect. This formula is independent of
the topology of the universe and is valid in the limit of an
infinitely thin last scattering surface and in absence of
reionization. In order to go beyond these approximations, one should
replace the quantities evaluated at the LSS, at $r \hat{\bf n}$, by
quantities averaged on spheres of various radii, weighted by the
probability for an electron to have experienced its last scattering at
the corresponding epoch~\cite{HuCMB}. Such corrections are however not
relevant for our purpose here.
 
In a multiply connected space, any two comoving points of space are
joined by more than one geodesic.  This characterizes the imprint of
topology on cosmology. The immediate consequence is that the observed
sky may show multiple images of a radiating source. Geometrically, the
observational space identifies with the portion of the covering space
(i.e. the corresponding simply connected manifold) which lies inside
the horizon sphere.  Multiple images (also called topological, or
ghost images) of a given source are related by discrete isometries
belonging to the holonomy group. The actions of these holonomies tile
the observational space into identical cells which are copies of a
fundamental domain. It results the most intuitive way to detect a
multiply connected topology: to identify the multiple images of the
same celestial object.  This applies to faraway cosmological sources
such as galaxy clusters~\cite{lelalu96} as well as spots in the CMB,
i.e., different points of the LSS which would correspond to a single
point of physical space.

When the fundamental domain is smaller than the size of the LSS (at
least along one direction), the multiple images of the LSS so
generated in the covering space intersect themselves.  In this case,
the Sachs-Wolfe contributions to the CMB radiation (which do not
depend on the direction of observation) are strictly identical for
homologous points of the LSS. This is at the basis of the now popular
``circles-in-the-sky'' method, since the intersection of two copies of
the LSS sphere is a circle.  Accordingly, a multiply connected
topology may be characterized by temperature correlations between
pairs of specific matched circles~\cite{topocmb}.

However, even in an ideal situation where noise removal would be
perfect, such a correlation is not perfect.  It is exact for the
Sachs-Wolfe contribution (and even only on scales larger than the
width of the last scattering surface), but absent for the Doppler and
ISW contributions~\cite{flatspaces}. In particular, the ISW
contribution depends on the whole path between the emitting point of
the LSS and the observer. In a multiply connected space, two matched
points of the LSS are joined to the observer by two different paths.
Therefore, their corresponding ISW contributions differ significantly,
except possibly on the very largest scales. This blurs the temperature
correlations.

A simple estimator for the correlation between pairs of circles,
labelled by the two indices 1 and 2, is the \emph{circle comparison
  statistics} (hereafter CCS) introduced by \cite{topocmb} as:
\begin{equation}
S(\phi_*)\equiv \frac{2<\Theta_1(\pm
  \phi)~\Theta_2(\phi+\phi_*)>}{<[\Theta_1(\pm
  \phi)]^2+[\Theta_2(\phi+\phi_*)]^2>} 
\end{equation}
where $\Theta_i$ is the temperature fluctuation along circle $i =
1,2$, $\phi$ is the angle associated to the running point on the
circles, and $\phi_*$ is the possible relative phase between the two
circles. The average is done all along each circle, which means that
$<\,\,> = 1/2\pi\,\int_0^{2 \pi} \,\mathrm{d}\phi$. The $+$ sign in
this equation correspond to phased circles and the $-$ sign to
anti-phased circles arising in non-orientable topologies.  With this
set of notations, the CCS ranges in the interval $[-1, +1]$: circles
that are perfectly matched have $S = 1$, while uncorrelated circles
have a mean value of $S = 0$ and totally anti-correlated circles
$S=-1$.

Let us for instance estimate the expected correlation of the Doppler
contribution, which is the main term of blurring, in order to see the
relative magnitude of the perturbation induced by this term on the
total correlation. We will also suppose that the pairs of matched
circles remain phased. This property occurs for instance when they are
linked by a simple translation, like in the toroidal case. Without
loss of generality, let us fix the coordinate system such that the two
images of a point lie at colatitudes $\pm \alpha$ in directions $\hat
{\bf n}_1 = (\sin \alpha, 0, \cos\alpha)$, $\hat {\bf n}_2 = (\sin
\alpha, 0, - \cos\alpha)$. As seen in eq.~(\ref{TempFluc}), the
amplitudes of the Doppler terms in these two directions are
proportional to $\hat {\bf n}_1 . \hat {\bf N}$ and $\hat {\bf n}_2
. \hat {\bf N}$ respectively, where $-\hat {\bf N}$ represents the
normalized direction of the electron fluid velocity (the same for both
points according to our hypothesis). It results, between the two
Doppler contributions, a correlation:
\begin{equation}
S_D = \frac{<2\, \hat {\bf n}_1 . \hat {\bf N} \times \hat {\bf n}_2 .
\hat {\bf N}>}{<(\hat {\bf n}_1 . \hat {\bf N})^2 + (\hat {\bf n}_2 .
\hat {\bf N})^2>}.
\label{CD}
\end{equation}

In the above equation, the average correlation is computed by
averaging separately both the numerator and the denominator in the
fixed direction $\hat {\bf N}$. After simple algebra, this gives
\begin{equation}
\left<S_D\right> = - \cos (2\alpha) = \cos \beta ,
\end{equation}
where $\beta = \pi - 2 \alpha$ represents the angular separation of 
the two points.

\begin{figure}
\begin{center}
 \includegraphics[width=3.5in]{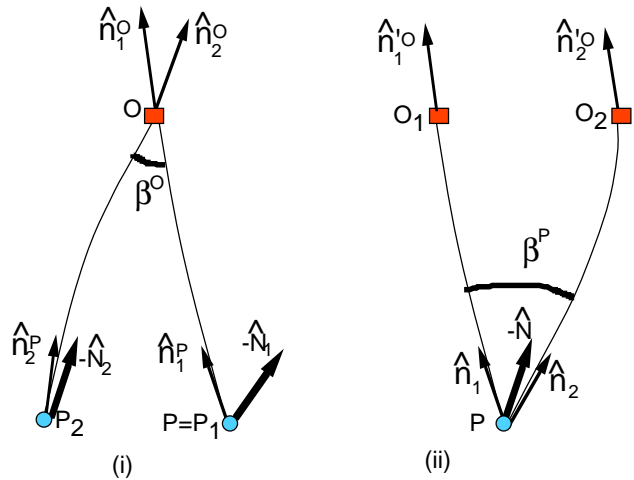}
\caption{Two points of view for describing the light-ray paths between
  a source point and the observer in a multiply connected space: (i)
  the observer $O$ sees two images $P_1$ and $P_2$ of the same point
  of the last scattering surface; (ii) two copies of the observer,
  $O_1$ and $O_2$, see the same point $P$ of the last scattering
  surface.}
\label{SpaceTimeWiew}
\end{center}
\end{figure}

Can we generalise this reasoning to an arbitrary geometry and
topology? In a multiconnected space, the observer $O$ sees multiple
images of an unique source point $P$ in real space (Fig.
\ref{SpaceTimeWiew}, (i)). Let $P_1$ and $P_2$ be two images of the
same point source $P$.  These images are of course related by $P_2 =
g(P_1)$, where $g$ is a holonomy transformation. The normalized
direction of the velocity of the electron fluid is represented by
$-\hat{\bf N_1}$ at $P_1$ and $-\hat{\bf N_2} = Dg( -\hat{\bf N_1})$
at $P_2$, where $Dg$ is the differential application corresponding to
$g$. Two light rays reach the observer at $O$ who can measure their
arrival directions $\hat{\bf n}^O_1$ and $\hat{\bf n}^O _2$. A strict
application of equation~(\ref{CD}) would imply calculating the scalar
products $\hat{\bf n}^O_1 . \hat{\bf N_1}$ and $\hat{\bf n}^O_2
. \hat{\bf N_2}$, hoping to express $S_D$ in terms of observed
quantities such as the angle $\beta^O$ between $\hat{\bf n}^O_1$ and
$\hat{\bf n}^O_2$ (it is well defined as $\cos \beta^O \equiv \hat{\bf
n}^O_1 . \hat{\bf n}^O _2$). But these scalar products are undefined
because the vectors involved are not attached to the same point. Thus
$S_D$ must be better defined.

To generalise the reasoning to an arbitrary geometry and topology, we
choose to adopt a dual point of view (Fig. \ref{SpaceTimeWiew},
(ii)). The point source $P$, with velocity direction $-\hat{\bf N}$,
emits light to two copies $O_1=O$ and $O_2 = g^{-1}(O_1)$ of the
observer. These two homologous observers see the source respectively
in the directions $\hat{\bf n}^{\prime O}_1=\hat{\bf n}^O_1$ and
$\hat{\bf n}^{\prime O}_2$, defined at $O_1$ and $O_2$ respectively.
These two directions cannot be compared, nor their scalar product
formed, since they correspond to vectors attached to different points
of the covering space. However, the two light rays reaching $O_1$ and
$O_2$ depart from the same point $P$ following spatial directions
$\hat {\bf n}_1$ and $\hat {\bf n} _2$.  This allows us to apply
equation~(\ref{CD}) to estimate $S_D$ as above. Averaging both the
numerator and the denominator over $\hat{\bf N}$ gives
$\left<S_D\right> = \cos \beta^P$ where $\beta^P$ defined by $\cos
\beta^P \equiv \hat{\bf n}_1 . \hat{\bf n}_2$ is the angle under which
the source ``sees" the two copies of the observer.

$\beta^P$ is not an observable quantity. There is in principle no
difficulty to convert $\beta^P$ into $\beta^O$, but this is a tedious
calculation, which requires the exact knowledge of the geometry of
space and the explicit expression of the holonomy transformation $g$,
and of its differential application $Dg$. This requires to estimate
the parallel transport of the vectors $\hat{\bf n}_1$ and $\hat{\bf
n}_2$ along the null geodesics (the light rays), as well as along the
trajectory of the holonomy transformation.  However, the situation is
greatly simplified when space is flat: then $Dg = \mathrm{Id}$ for a
translation, and $-\mathrm{Id}$ for a translation with space
inversion.  In this case, we have respectively $\cos \beta^P = \cos
\beta^O$ and $\cos \beta^P = -\cos \beta^O$.

Unsurprisingly, two antipodal points are fully anticorrelated since
they correspond to a Doppler effect seen from opposite directions,
whereas the limit where the two points are in the same direction gives
a fully correlated Doppler term. After a quick reminder about
polarization in the next section, we shall perform a similar analysis
for polarization.

\section{A quick reminder of CMB polarization}
\label{Secqr}

We recall that the propagation of a density wave in an anisotropic
plasma induces a linear polarization of the CMB~\cite{Rees68}. Each
photon is polarized when scattered off by an electron of the
photon-baryon plasma. The isotropic superposition of photons in the
CMB destroys on the average the polarization. However, the presence of
a local quadrupole in the photon fluid induces a weak residual
polarization in the direction orthogonal to the plane defined by the
quadrupole. The amount of \emph{observed} polarization depends on the
orientation of the observer with respect to the quadrupole, in analogy
with the Doppler term dependence on the relative angle between the
line of sight and the fluid velocity.  The perfect fluid approximation
for the photons would imply the absence of quadrupole. But, during the
decoupling, the perfect fluid approximation is broken by the dramatic
decrease of the scattering rate, as the free electrons become bound to
the atomic nuclei. This generates a linear polarization of the CMB
temperature. This prediction was only recently
verified~\cite{detpola}, because of the low amount of polarization
expected.

To analyze the generation of polarization, we may consider a density
wave, which generates density gradients in the photon fluid, in the
direction of its wavevector.  When the photons propagate, these
density gradients generate a local dipole in the photon fluid: in a
region initially richer in photons to its left than to its right,
appears a dipole oriented from left to right. It results a dipole
distribution, whose direction and intensity remain constant on a given
waveplane, but vary along the orthogonal direction. A similar
reasoning shows that this dipole gradient generates a local
quadrupole, and so on. This is just a rewording of the Boltzmann
equation, which couples the $\ell$-th multipole of the radiation to
the $(\ell-1)^{th}$ and $(\ell+1)^{th}$ multipoles.

Given such a local quadrupole, the scattering of photons on free
electrons generates a polarization as explained above.  Since the
orientation of the dipole and quadrupole are determined uniquely by
the direction of the initial density field, the generated polarization
also depends on the angle between the line of sight and the quadrupole
direction.

For convenience, the usual linear $Q$ and $U$ Stokes parameters,
measured in each direction of the microwave sky (and assuming some
rather arbitrary choice of the axis of the polarizer), are split into
``electric'' and ``magnetic'' parts, through a non local
transformation. These names reflect the intuitive properties of the
corresponding patterns with respect to parity transformation. For
symmetry reasons, the density fluctuations create only $E$-modes
(i.e. curl-free) polarization patterns on the sky. Although one
expects that tensor modes (gravitational waves) also contribute to the
CMB fluctuations, the scalar (density) modes are expected to be
dominant~\cite{revcmb}. Thus, we assume hereafter that polarization is
due to scalar modes only, so that we only consider the angular
dependence of the amplitude of the scalar $E$ modes.

\section{Correlation in polarization maps}
\label{SecPol}

The polarization is estimated by a method similar to that of
Sec.~\ref{Seccorr}. The only difference lies in the angular dependence
of the scalar part of the polarization tensor.

The amplitude of the temperature fluctuation due to the Doppler term
is proportional to $\hat {\bf n} . \hat {\bf N} = \cos \theta$, where
$\theta$ is the angle between the fluid velocity direction $\hat {\bf
N}$ and the line of sight $\hat {\bf n}$.  Thus, the total angular
momentum method associates naturally the scalar part of this Doppler
term with the spherical harmonics $Y^0_1$. In a similar
way~\cite{HuWhite97}, one can associate the scalar part of the
polarization with the spin-weighted spherical harmonics $_{2}Y^0_2
\propto \sin^2\theta = 1 - (\hat{\bf n}.\hat{\bf N})^2$. Although far
less intuitive, this can be understood from purely geometrical
considerations as indicated in \cite{HuWhite97}. This allows us to
compute the Doppler contribution of the expected correlation $S_D^E$
of the scalar $E$ mode of polarization for the two points, by a method
similar as above: one simply has to replace in (\ref{CD}) the angular
dependence $\hat{\bf n}.\hat{\bf N} = \cos \theta$ by $1 - (\hat{\bf
n}.\hat{\bf N})^2 = \sin^2 \theta$. Therefore, one obtains:
\begin{equation}
S_D^E = \frac{<2\, (1 - (\hat {\bf n}_1 . \hat {\bf N})^2) \times (1 -
(\hat {\bf n}_2 . \hat {\bf N})^2)>}{<[1 - (\hat {\bf n}_1 . \hat {\bf
N})^2]^2 + [1 - (\hat {\bf n}_2 . \hat {\bf N})^2]^2>} ,
\end{equation}
and, after averaging,  
\begin{equation}
\label{ces}
S_D^E = 1 - \sin^2\alpha\cos^2\alpha
      = 1 - \frac{1}{4} \sin^2 \beta
\end{equation}
Thus, in the worst case (points in orthogonal directions), the
correlation level still remains above $75\%$, whatever the value of
$\beta$, and increases toward 1 for antipodal points. But $\beta$ is
the angle between the two copies of the observer as seen from the
point source and is different from the observed angle $\beta^O$
between the copies of the source as seen by the observer. Thus,
formula (\ref{ces}) could not directly gives us the observed
variations of $S_D^E$ but indicates that the blurring due to the
Doppler effect is always weak, more precisely that $S_D^E(\beta^O)\ge
0.75$, for all $\beta^O$ as it is true for all $\beta$.

\section{Simulated maps and results}
\label{SecSim}

In order to illustrate the efficiency of the method, we apply it, as
an example, to two simple cases of multiconnected flat spaces: the
3-torus and the sixth-turn space~\cite{flatspaces}. We consider a flat
$\Lambda$CDM model with density parameters $\Omega_\Lambda = 0.7$,
$\Omega_{\rm cdm} = 0.3$, and a Hubble constant $H_0 = 72 \; {\rm
km}\;{\rm s}^{-1}\;{\rm Mpc}^{-1}$; we neglect reionization. We used
our CMB code to compute maps of the scalar CMB polarization $E$, with
the method described in~\cite{flatspaces}. The only difference is that
the quantities $\Theta_\ell (k, \eta)$ are replaced by their
polarization counterparts $E_\ell (k, \eta)$.

\begin{figure}


\centerline{\psfig{file=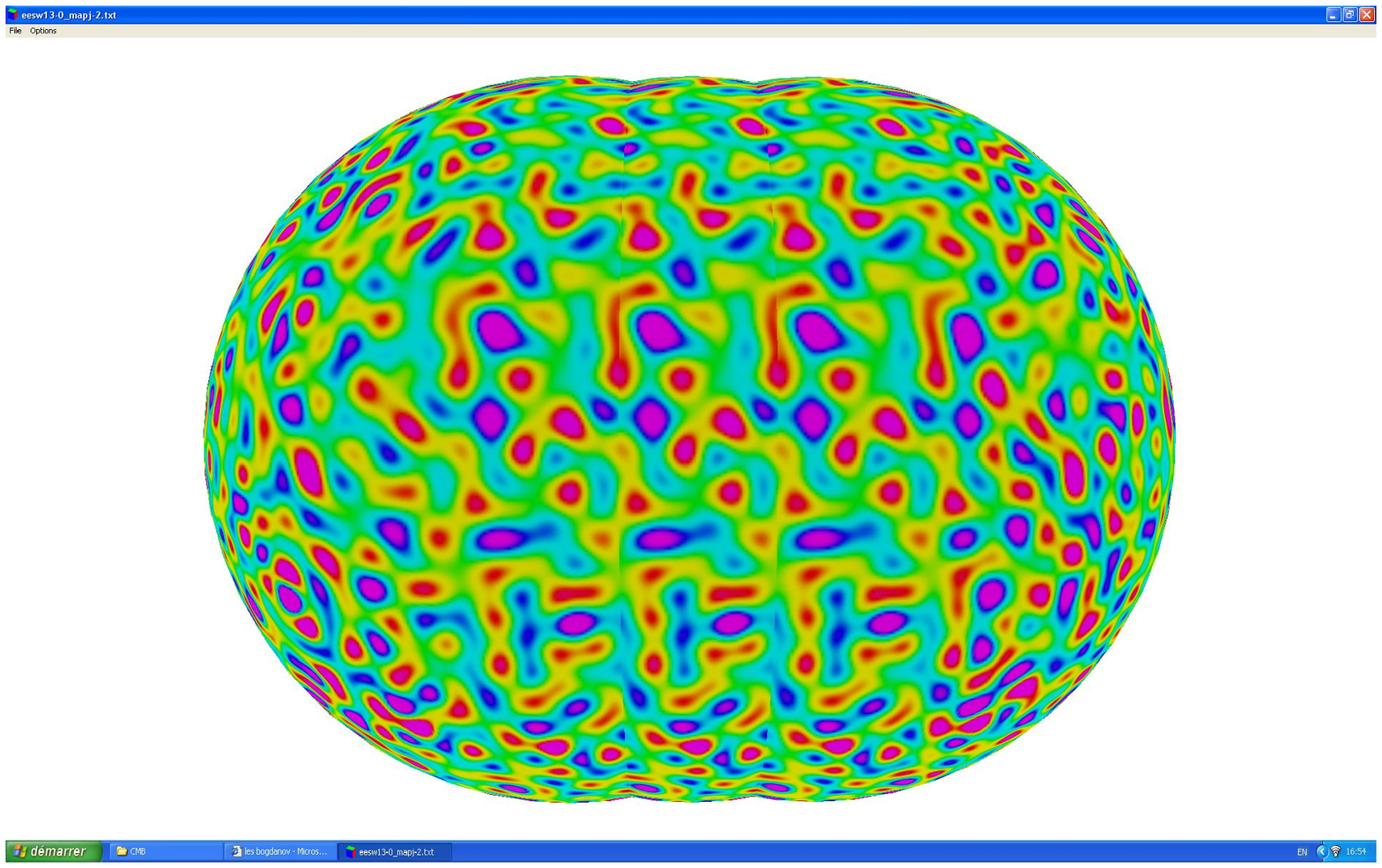,angle=0,width=3.5in}
            \psfig{file=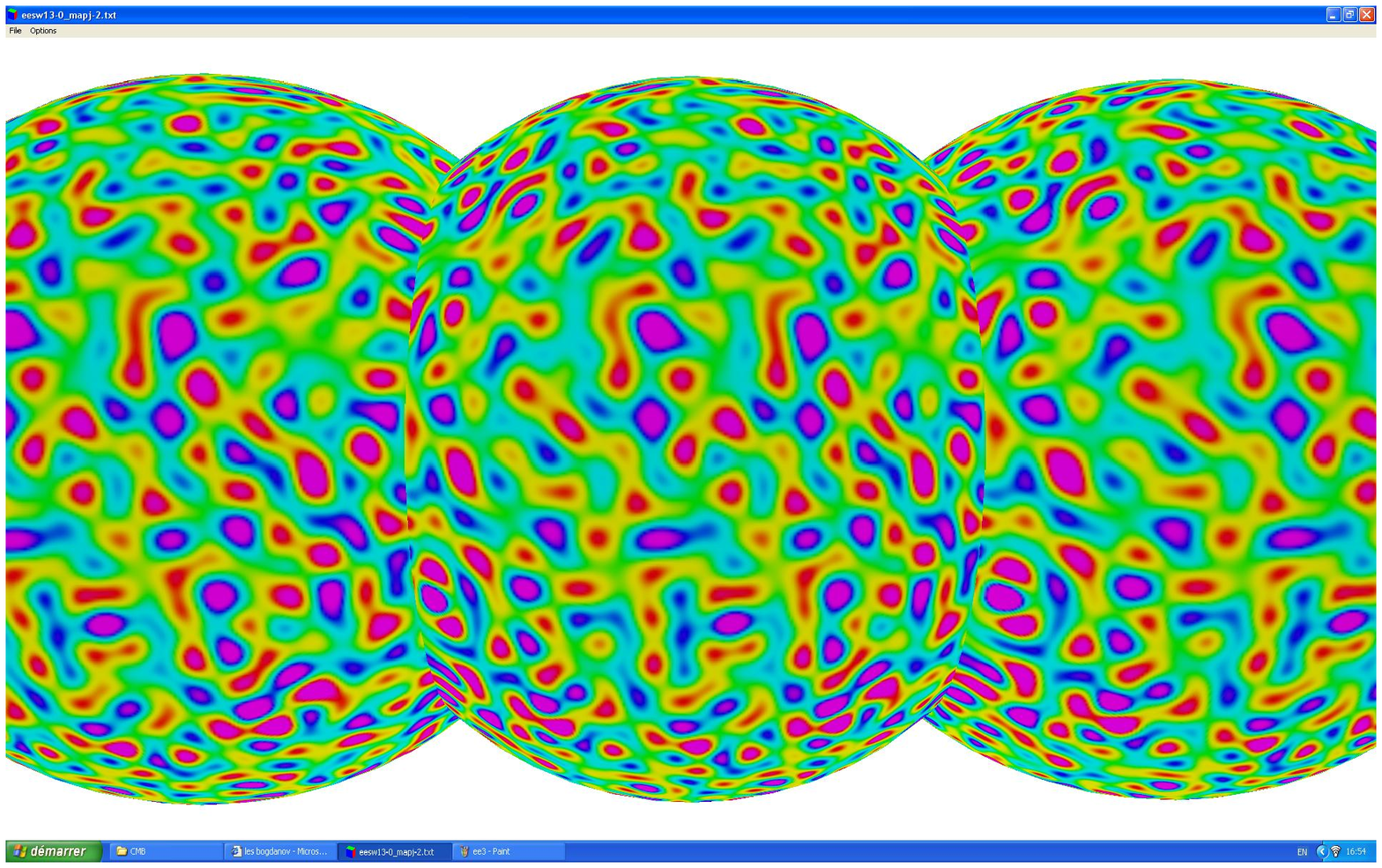,angle=0,width=3.5in}}
\centerline{\psfig{file=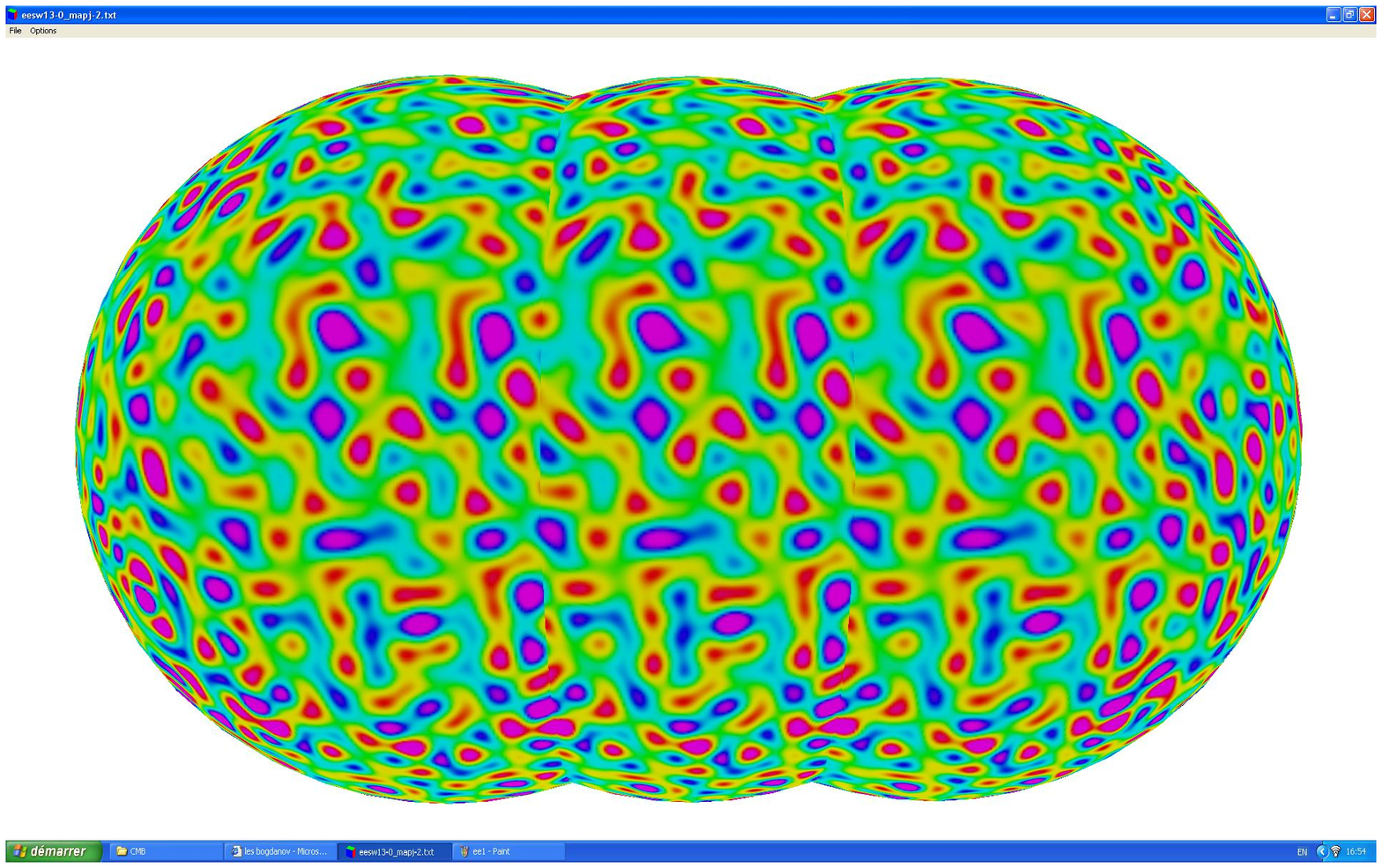,angle=0,width=3.5in}
          \psfig{file=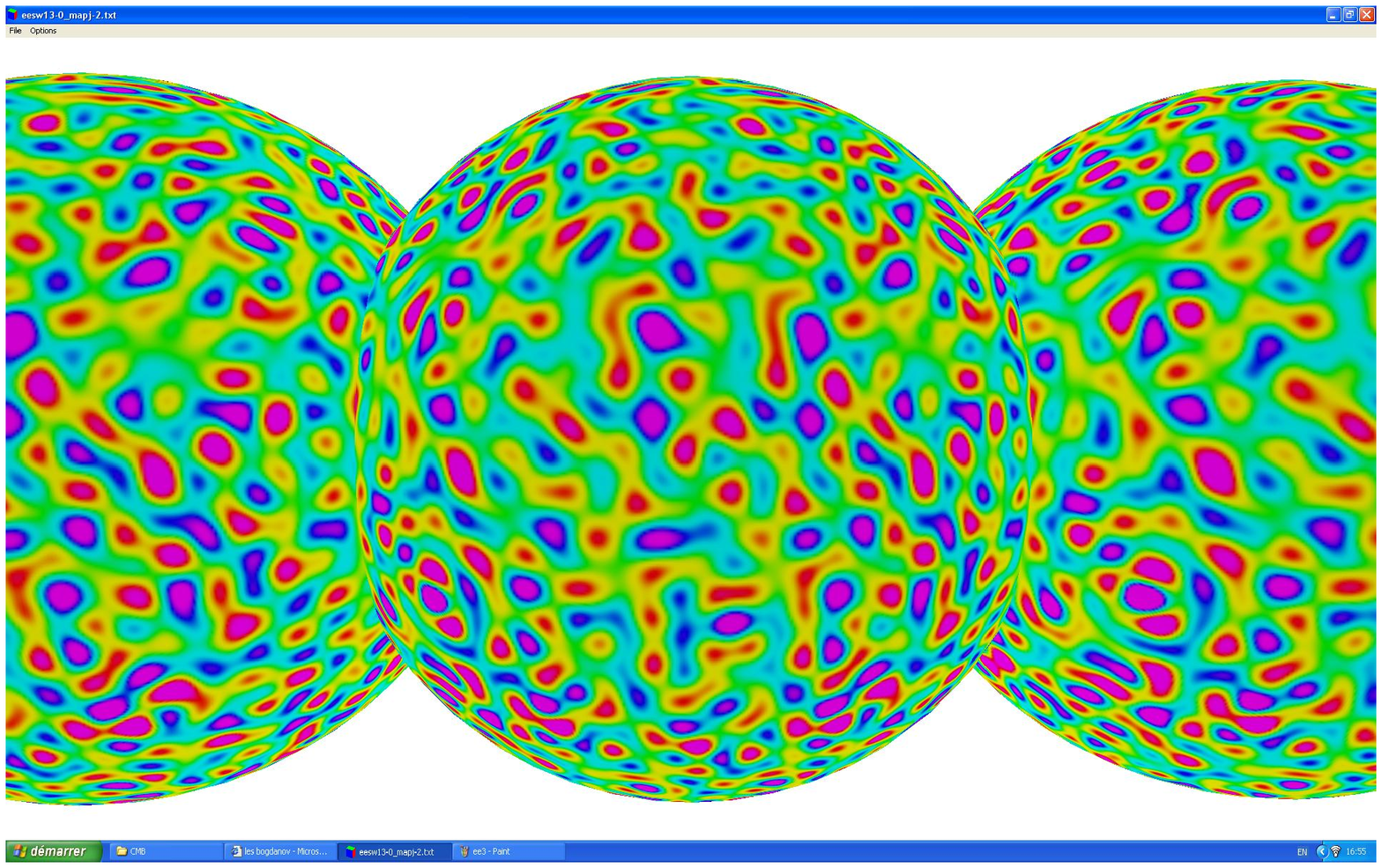,angle=0,width=3.5in}}
\centerline{\psfig{file=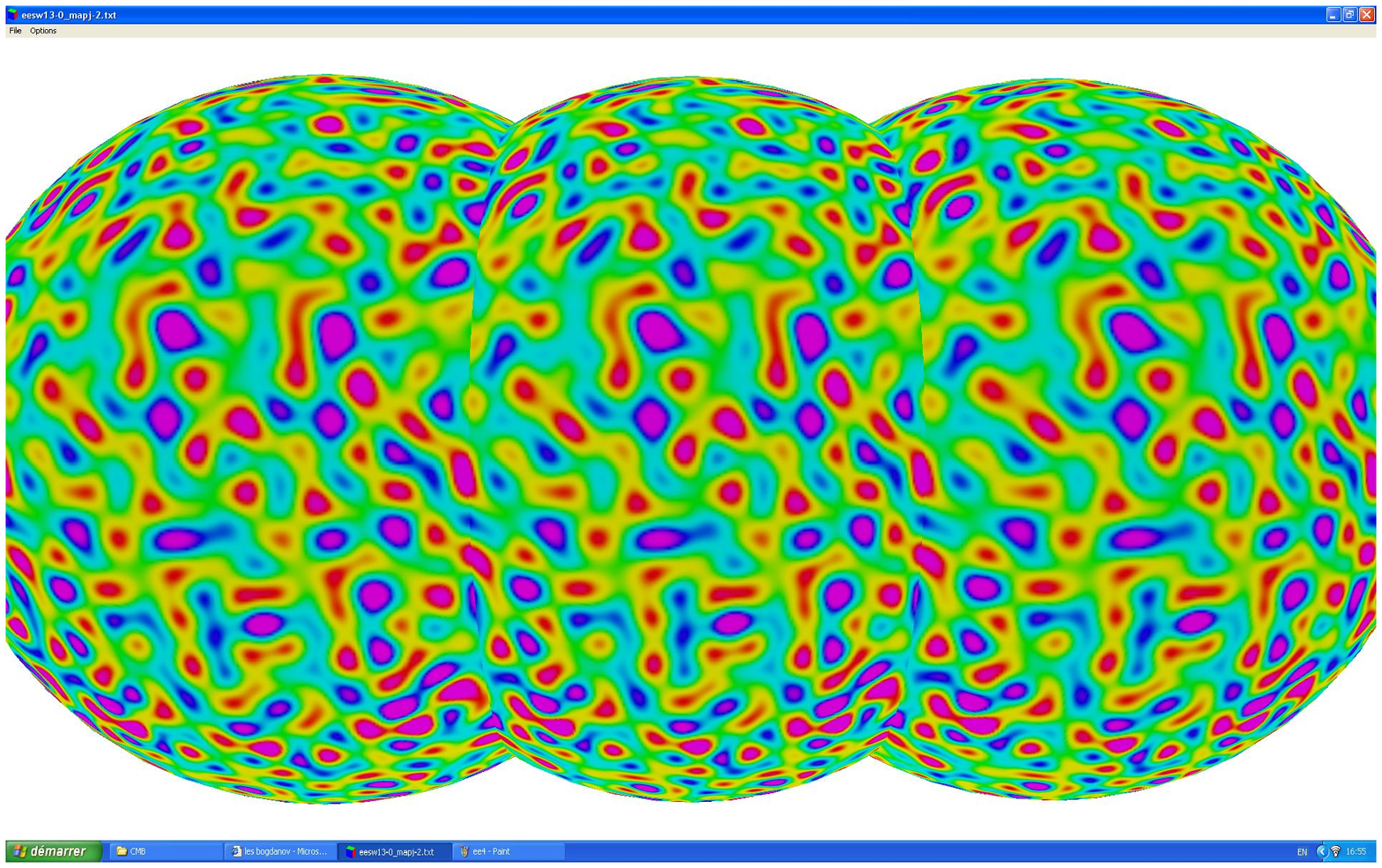,angle=0,width=3.5in}
            \psfig{file=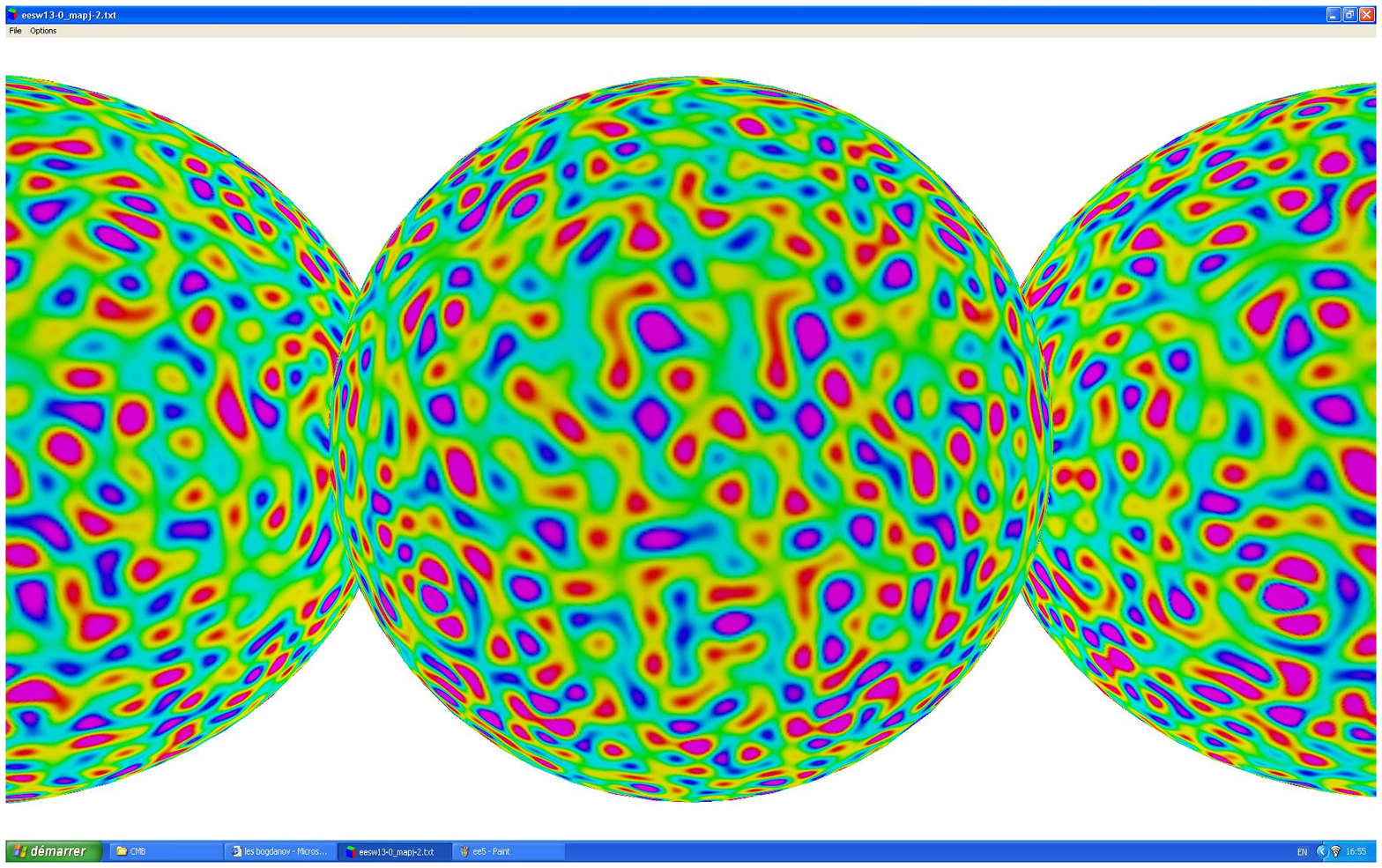,angle=0,width=3.5in}}
\caption{The CMB scalar $E$ polarization anisotropies in a flattened
toroidal universe. We consider here a 3--torus with two lengths equal
to the diameter of the LSS and one six times smaller (corresponding to
the horizontal direction of the figure). We show three copies of the
LSS along the small length, whose intersections correspond to the
matched circles.  We consider a small number of modes. No filtering
has been applied to the map. Therefore, all the angular scales to
which the modes contributes are faithfully represented here. The
matching is very good for the largest nearby circles. It then
decreases, and increases again for the smallest circles, as follows
from Eq.~(\ref{ces}). It is the worst case for circles of radius $45$
degrees, as seen in the bottom left and upper right configurations.}
\label{figee}
\end{figure}

\begin{figure}


 \centerline{\psfig{file=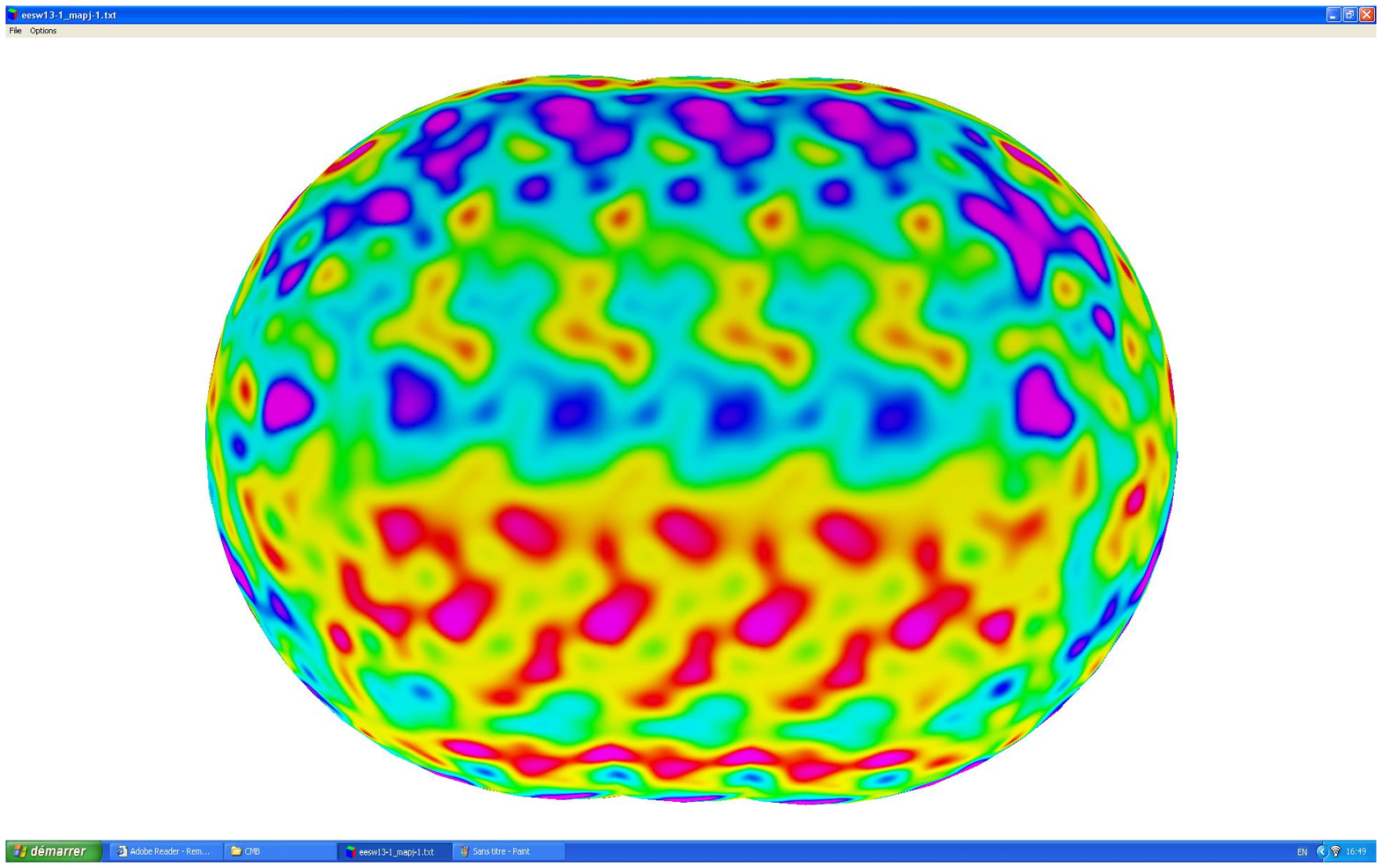,angle=0,width=3.5in}
            \psfig{file=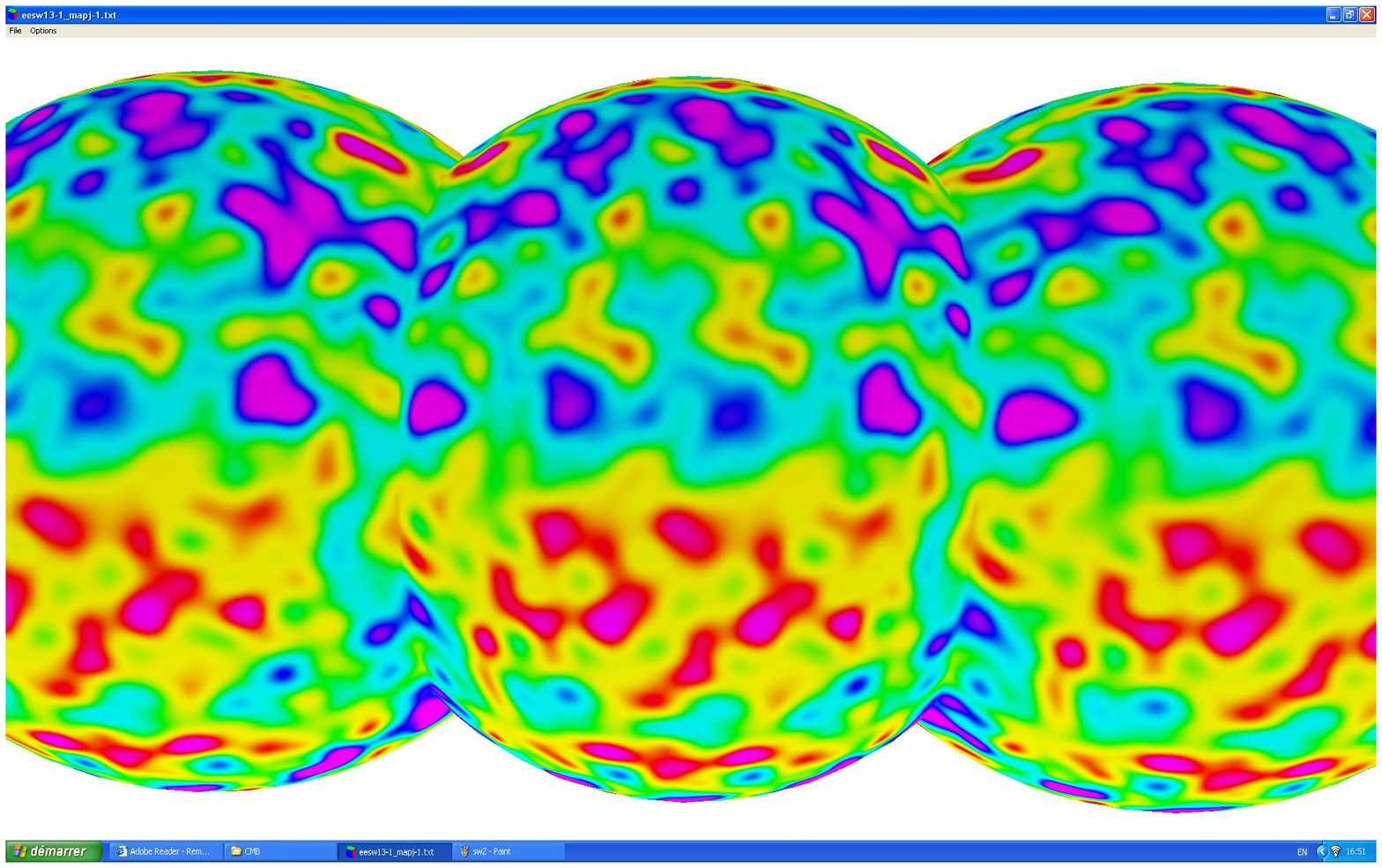,angle=0,width=3.5in}}
 \centerline{\psfig{file=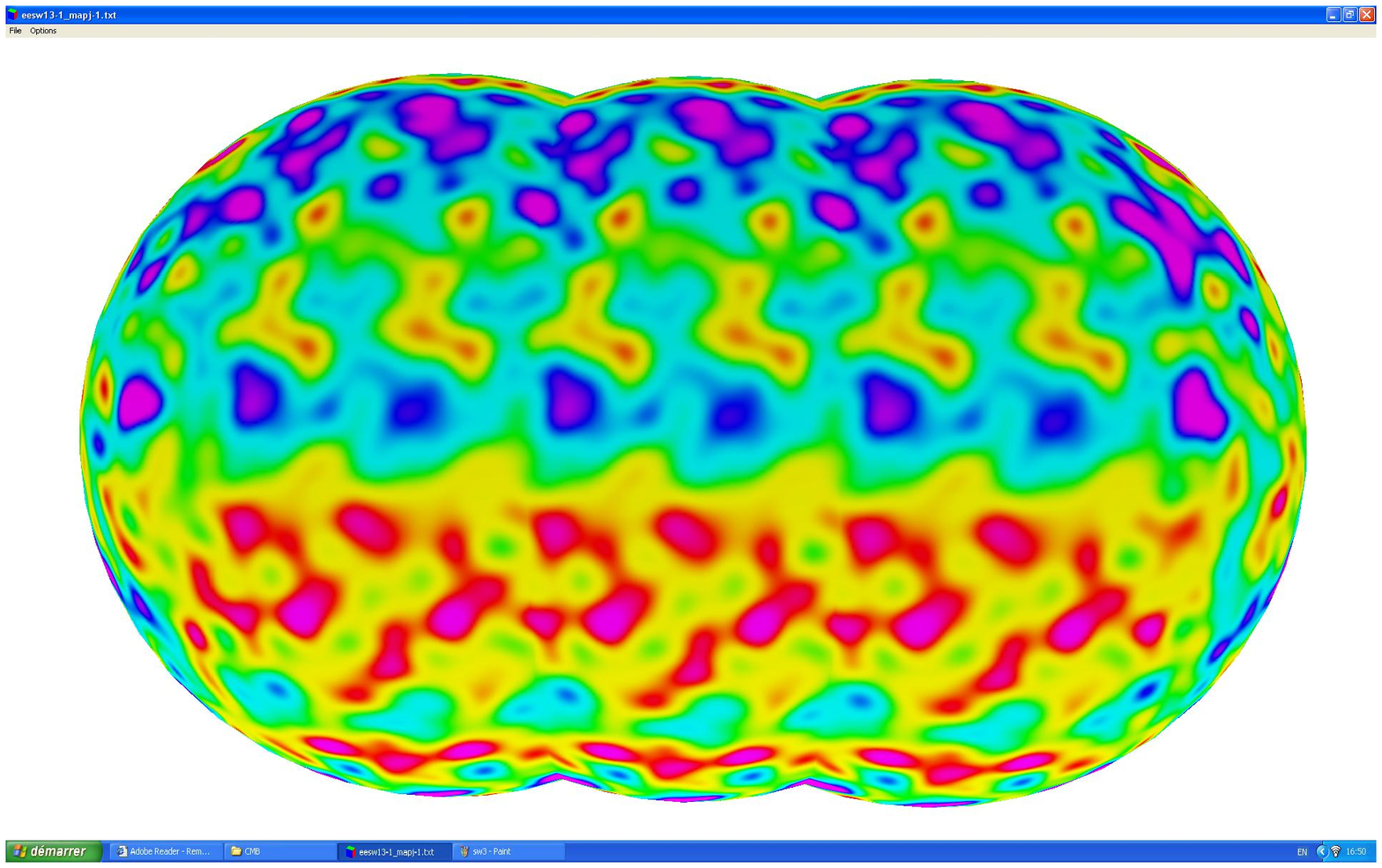,angle=0,width=3.5in}
             \psfig{file=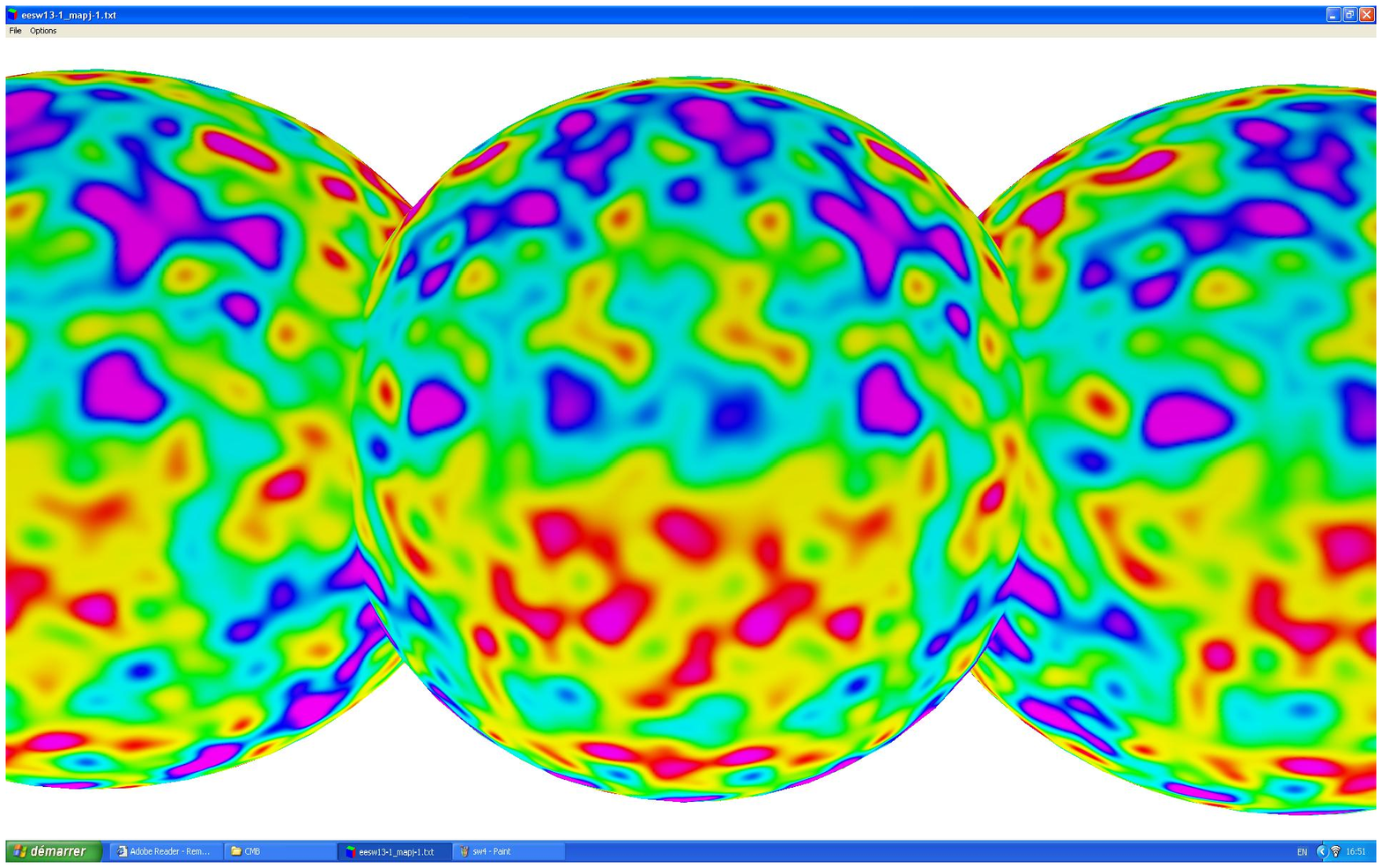,angle=0,width=3.5in}}
 \centerline{\psfig{file=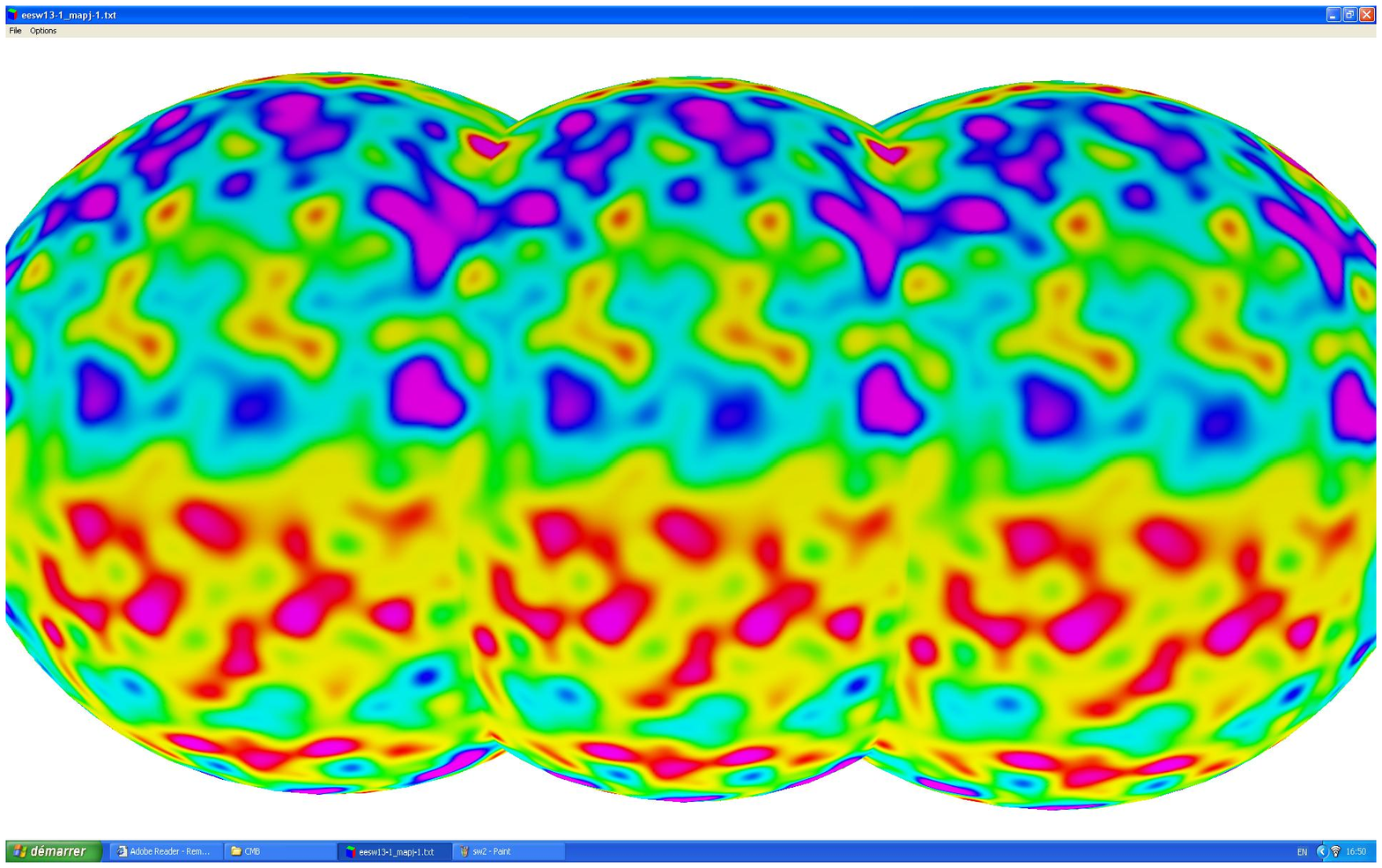,angle=0,width=3.5in}
            \psfig{file=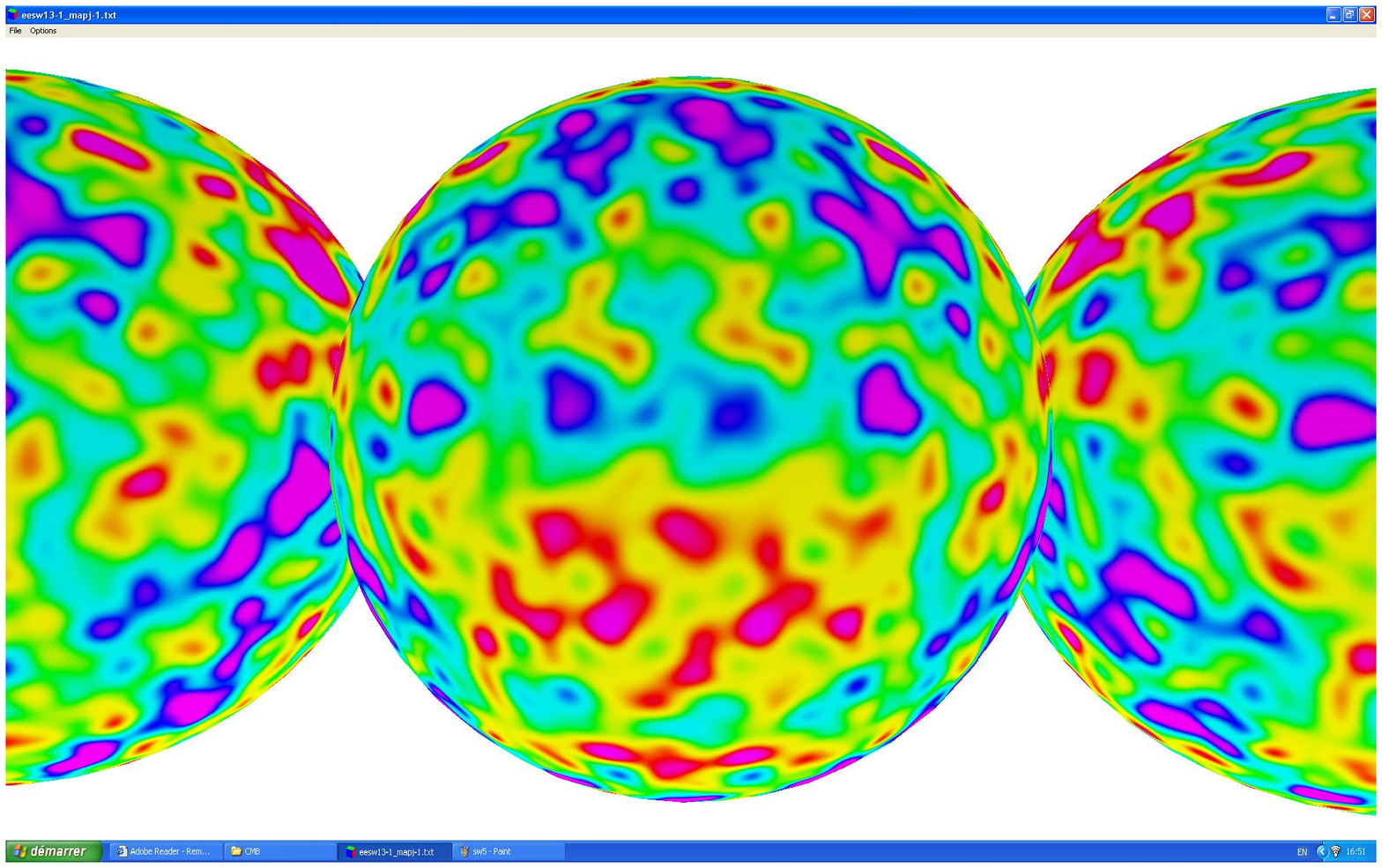,angle=0,width=3.5in}}
\caption{Same as Fig.~\ref{figee}, but for the Sachs-Wolfe
contribution. Since only the Sachs-Wolfe contribution is considered
here, the matching between the circles is almost perfect, regardless
of the angular size or distance of the circles. Note that the map
seems to be at a lower resolution than the polarization map since the
former has a relatively flat $\ell (\ell + 1) C_\ell$ spectrum,
whereas the latter has a fast growing spectrum because no large scale
polarization is present.}
\label{figsw}
\end{figure}

For the 3-torus, low resolution maps are presented in
Fig.~\ref{figee}. The main result is that the polarization
fluctuations are always positively correlated, although not as
perfectly as for the (academic) pure SW temperature map, as shown
after in Fig.~\ref{figsw}.

\begin{figure}


\centerline{\psfig{file=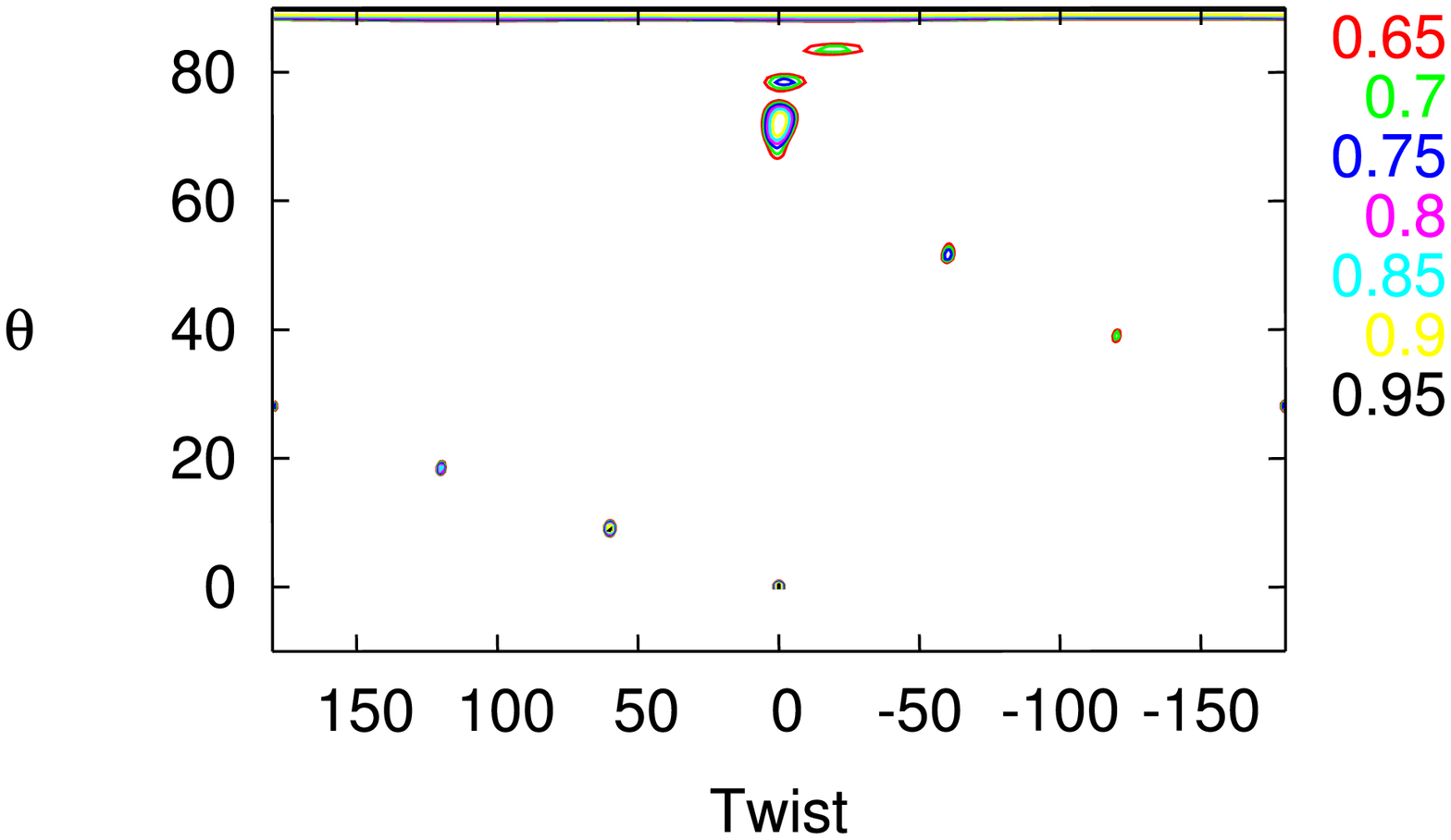,bbllx=135pt,bblly=150pt,bburx=500pt,bbury=630pt,angle=270,width=3.2in}}
 
 \centerline{\psfig{file=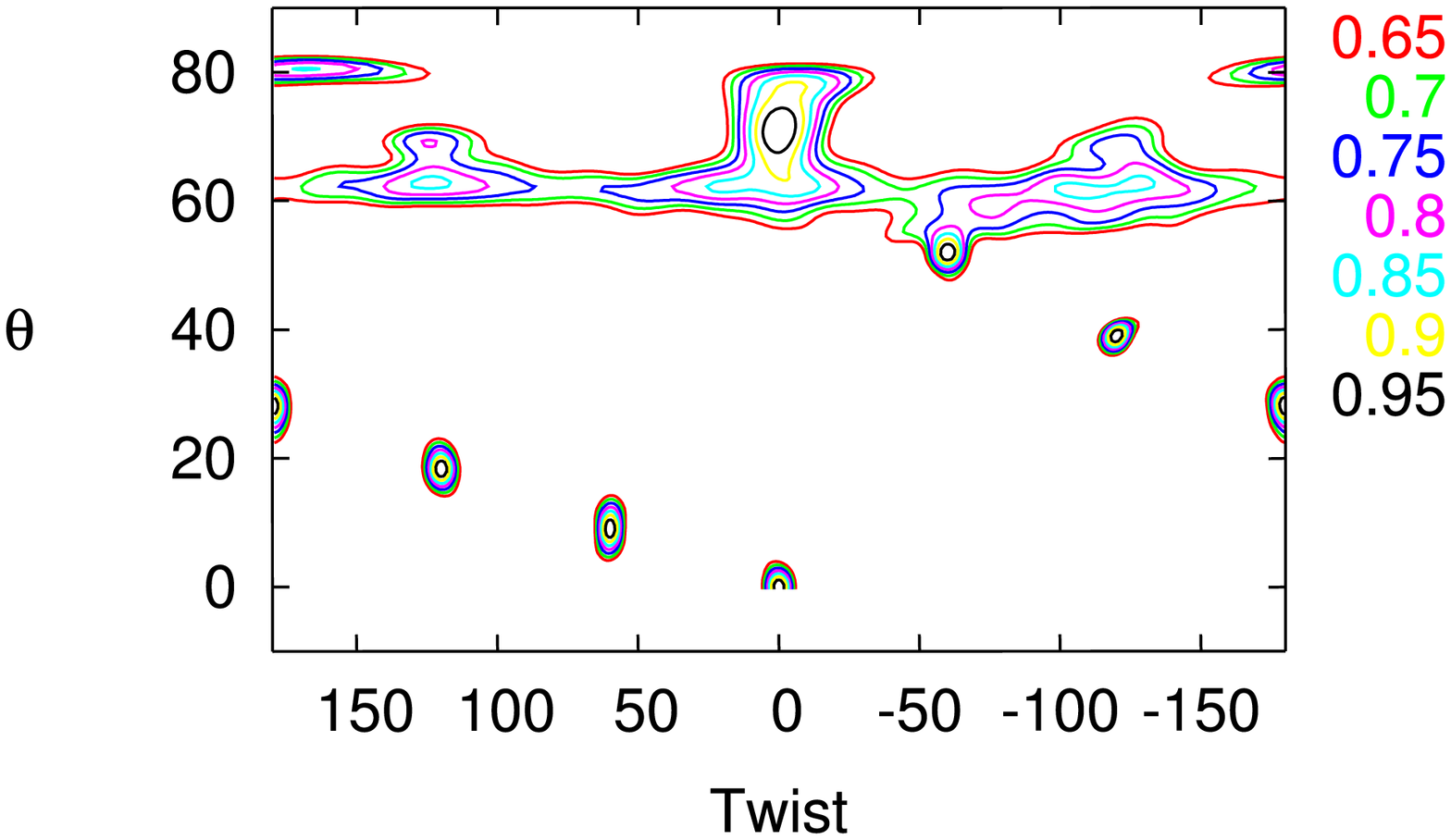,bbllx=135pt,bblly=150pt,bburx=500pt,bbury=630pt,angle=270,width=3.2in}}

 \centerline{\psfig{file=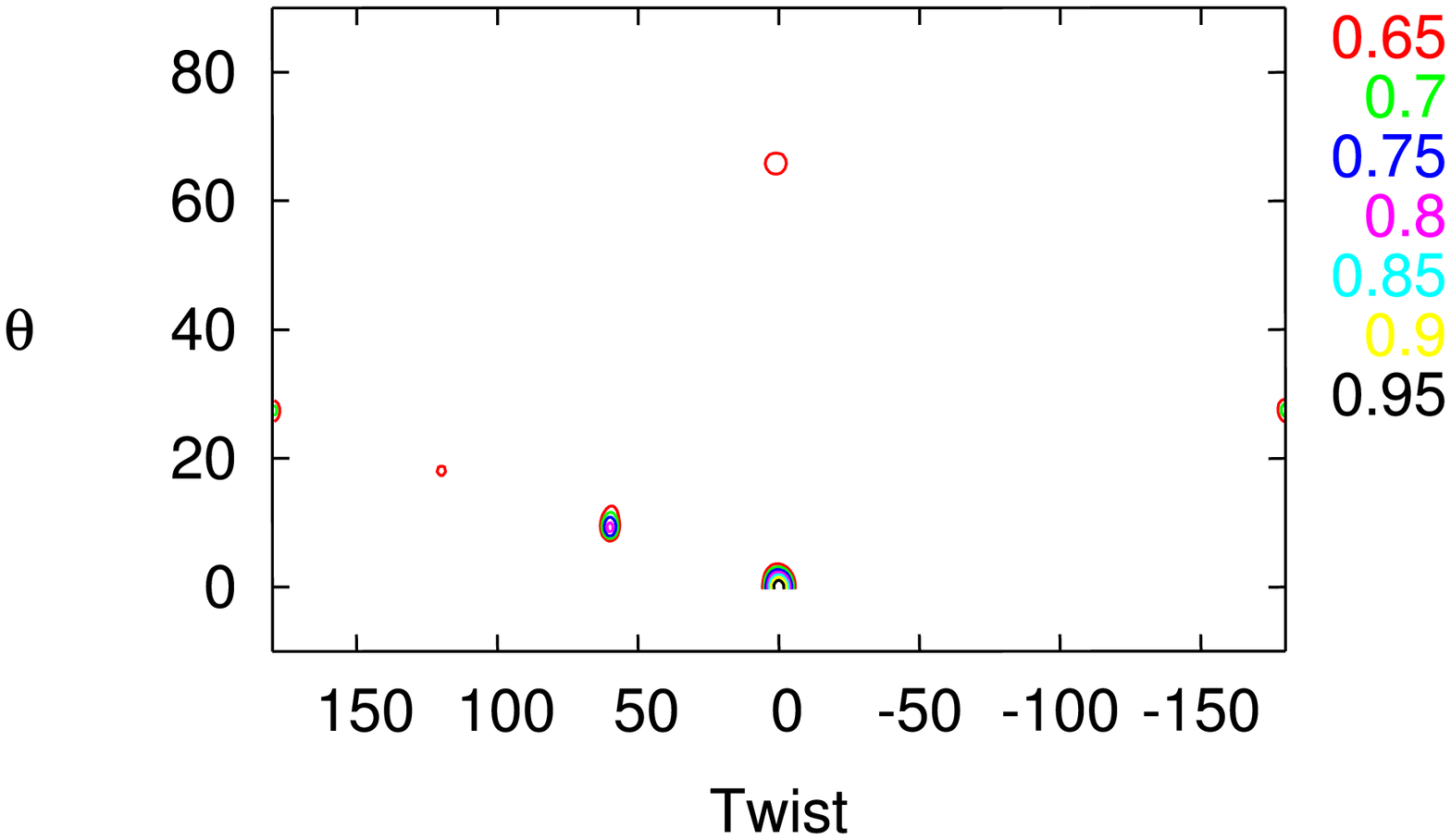,bbllx=135pt,bblly=150pt,bburx=500pt,bbury=630pt,angle=270,width=3.2in}}

\caption{Comparison of the circle matching for a scalar polarization
$E$ map and a temperature map. In these plots we compute the
correlation between circles of latitudes $\pm \theta$, with a possible
twist between them in a sixth-turn space. Given the small correlation
length of CMB anisotropies in any realistic cosmological model, the
correlation between circles is expected to be negligible except when
the circles are very close (around $\theta = 0$ with a negligible
twist), or for the values $\theta_n$ and $\varphi_n$ given in
Eq.~(\ref{thetan})--(\ref{phin}) given by the topology we consider
here. The first plot shows the correlation for a polarization map,
whereas the next one shows the correlation for the pure Sachs-Wolfe
effect, then the full temperature map for the same density
field. Although the pure (but unobservable) Sachs-Wolfe contribution
always gives as expected a 100\% correlation (up to pixellization
effects) for the matching circles, it is no longer the case for the
other maps. Note that the correlation in the polarization map starts
from 100\% for untwisted equatorial circles, then decreases and
reaches a minimum for circles of latitude $\sim 40$~degrees, and then
increases again for smaller circles, as expected. This pattern is
always present in polarization maps. For comparison, a temperature map
may not show such a regular pattern, in particular at low or
intermediate angular scales such as in these maps, where either the
ISW or Doppler terms blur the correlations.  }
\label{fig_circles}
\end{figure}

For the sixth-turn space we calculate the correlations between circles
on various simulated CMB maps. For this space, the fundamental domain
is an hexagonal prism. But along the direction of the prism, the
different copies are not deduced from a translation (as it is the case
for a torus), but from a screw motion defined as a translation
composed with a $\pi/3$ rotation along the vertical axis (here
corresponding to the direction of the prism and to the vertical
direction in the maps). The observer sitting at the center of the
prism expects to see pairs of matched circles at opposite
latitudes. In each pair the circles are relatively twisted by an angle
$n \pi / 3$, where $n$ is the number of screw motions one has to
perform to go from one circle to the other. The prism height $L$ is
taken equal to one Hubble radius. This implies a radius of the
observable universe $R_{\rm LSS} \sim 3.246$ Hubble radii. Therefore,
we expect a series of matched circles at latitudes

\begin{equation}
\label{thetan}
\theta_n = \pm \arccos\left(n \frac{L}{2 R_{\rm LSS}}\right) ,
\end{equation}
with $1 \leq n \leq 6$. The twist between the circles of latitude
$\pm \theta_n$ is  
\begin{equation}
\label{phin}
\varphi_n = n \frac{\pi}{3} .
\end{equation}
Since for a sixth-turn space, both the direction of observation and
the angular dependence of the emitting region are subject to the same
screw motions, the above formula~(\ref{ces}) remains valid.
Fig.~\ref{fig_circles} compares the circle matching for a scalar
polarization $E$ map and a temperature map. The dependence of the
matching of the circles as a function of their distance is more
straightforward.

\section{Possible sources of blurring of the circles}
\label{SecBlur}

The nice result of Eq.~(\ref{ces}) will however not correspond to what
can be observed, at least for four reasons:
\begin{enumerate}

\item The relevance of Eq.~(\ref{ces}) relies on the assumption that
all the polarization is generated at the last scattering epoch, and
does not take into account any equivalent of the ISW effect for the
polarization. However, the duration of the last scattering epoch is
short only in the absence of reionization. Since the last WMAP
results~\cite{reswmap} suggest that a significant amount of
reionization took place, maybe as early as $z \sim 15$, one expects
that some part of the observed CMB radiation was scattered long after
recombination. This should induce a bump in the observed polarization
spectrum at large angular scale (as well as a bump in the
cross-correlated $TE$ spectrum, as already claimed by the WMAP
team). In this case, the angular scales at which reionization
contributes should blur the circles. However, the angular power
spectrum is strongly increasing in this region. Thus, at intermediate
($\ell \sim 60$) angular scales, the low $\ell$ contribution of
reionization should become negligible, as compared to that of the
recombination epoch.

\item Equivalently, gravitational lensing is expected to contribute to
the CMB polarization anisotropies at large $\ell$. The relevant range
is however expected to be sufficiently different (large angular scales
for reionization, small ones for lensing), as to preserve a limited
range of $\ell$ where both reionization and lensing are negligible.

\item We have assumed a perfect signal-to-noise ratio in the
data. This is certainly the most objectable hypothesis done here,
since at present very little is known on how to extract as accurately
as possible foreground polarized emission from CMB observations. A
careful analysis of the expected signal-to-noise ratio of WMAP and
Planck mission polarization maps and its imprint on the detectability
of the topology of the universe should be performed and is left for
future studies.

\item Finally, Eq.~(\ref{ces}) was calculated under the assumption
that the whole set of angular scales to which a plane wave (say)
contributes was observed. In practice, an angular filtering is applied
to a map, in particular to account for the instrument resolution. A
single mode with comoving wavenumber $k$ mostly contributes to the
angular scale $\ell$ such that
\begin{equation}
\label{kl}
\ell~ \sim k \eta_{\rm LSS}, 
\end{equation}
where $\eta_{\rm LSS}$ is the comoving distance of the last scattering
sphere. However, this is only approximate, and a plane wave usually
contributes also, although with less intensity, to all the angular
scales larger than this value (see ~\cite{HuCMB}). As a consequence, a
map of resolution $\ell_{\rm max}$ exhibits in principle perfectly the
correlations due to modes with $k \lesssim k_{\rm max} = \ell_{\rm
max} / \eta_{\rm LSS}$, but also incompletely exhibits the
contributions from larger wavenumbers. The consequence is a decrease
of the expected correlation. To avoid this problem, the best way is to
choose a value of $\ell_{\rm max}$ which corresponds to a decreasing
part of the spectrum, making the contribution of higher wavenumbers $k
> k_{\rm max}$ as small as possible. For this reason, in a pure
Sachs-Wolfe map, the correlation obtained when the resolution
corresponds to the decrease of the first peak ($\ell_{\rm max} \sim
300 - 400$) is better than the correlation taken before the first peak
maximum ($\ell_{\rm max} \sim 150$). It happens that the most detailed
circle searches in CMB temperature map~\cite{circglenn, aurich2005}
were precisely performed at angular resolution $\ell \sim 400 - 500$,
so that this possibly annoying problem was evaded.

In other words, there is the necessity to convolve the
three-dimensional Fourier spectrum with some window function in order
to project it on the celestial sphere and obtain the $C_\ell$.  This
window function is not infinitely narrow as was done in the
approximation of Eq.~(\ref{kl}). However, the window function
associated to the scalar $E$ mode of polarization is in fact far more
peaked than that of the Sachs-Wolfe contribution to temperature
anisotropies, mainly because of the intrinsic angular dependence of
the polarization~\cite{HuCMB}. Therefore, this effect does seem to be
problematic for the imprint of topology in polarization maps.

\end{enumerate}

As we explained, among these four problems, the third one is certainly
the most serious. It is possible that even the Planck mission
sensitivity to polarization will not increase the signal-to-noise
ratio sufficiently enough to allow to search for the topology of the
universe in that way.  Future polarization designed experiments,
however, should be able to do the job. The problem addressed in this
paper represents an example of the required sensitivity that might be
imposed when designing next generation CMB experiments.

\section{Conclusion}

High precision experiments such as WMAP now give a clear outline of
the cosmological model of our universe. Despite this tremendous
success of modern cosmology, some old questions such as the shape of
space are still unanswered. In this paper, we have described a new
test to search for the topology of the universe using polarization
maps. This test is rather ambitious as it necessitates clean data on
the CMB polarization map.  Given the advantages of the polarization
maps which have been presented here, it appears highly likely that
future missions with sufficient good signal-to-noise ratio (Planck, or
some future polarization dedicated missions) will help bringing
definitive conclusions about the question of cosmic topology.

\acknowledgments

The authors wish to thank Simon Prunet, Glenn Starkmann and
Jean-Philippe Uzan for helpful discussions about CMB physics and
topology of the universe. Most of the simulations presented here were
performed on the MPOPM cluster of the Meudon Observatory computing
center. A.R.\ wishes to thank Prof.\ Yin for enlightening discussions
about floating point arithmetics.



\begin{thebibliography}{100}

\bibitem{revmeth} M.~Lachi\`eze-Rey \& J.-P.~Luminet, Phys.\ Rep.\
{\bf 254}, 135 (1995); J.~Levin, Phys.\ Rep.\ {\bf 365}, 251 (2002);
M.~J.~Reboucas, astro-ph/0504365 ; T.~Souradeep and A.~Hajian,
astro-ph/0502248.

\bibitem{revcmb} See for example A.~R.~Liddle \& D.~H.~Lyth, {\it
Cosmological Inflation and Large-Scale Structure}, Cambridge
University Press, Cambridge, England (2000).

\bibitem{cobe} G.~F.~Smoot {\it et al.}, \apj {\bf 396}, L1 (1992).

\bibitem{topocmb} N.~J.~Cornish, D.~N.~Spergel \& G.~D.~Starkman,
Class.\ Quant.\ Grav.\ {\bf 15}, 2657 (1998).

\bibitem{circglenn} N.~J.~Cornish, D.~N.~Spergel, G.~D.~Starkman \&
E.~Komatsu, \prl {\bf 92}, 201302 (2004).

\bibitem{circrouk} B.~Roukema {\it et al.} , Astron. Astrophys. {\bf
423}, 821 (2004).

\bibitem{lum03} J.-P.~Luminet, J.~Weeks, A.~Riazuelo, R.~Lehoucq \&
J.-P.~Uzan, Nature (London) {\bf 425}, 593 (2003).

\bibitem{flatspaces} A. Riazuelo {\it et al.}, \prd {\bf 69} 103518
(2004); A.~Riazuelo {\it et al.}, \prd {\bf 69}, 103514 (2004).

\bibitem{aurich04} R.~Aurich, S.~Lustig \& F.~Steiner,
astro-ph/0412569.

\bibitem{anomisocmb} H.~K.~Eriksen {\it et al.}, \apj {\bf 605}, 14
  (2004); Erratum {\it ibid.} {\bf 609}, 1198 (2004); C.~J.~Copi {\it
  et al.}, \prd {\bf 70}. 043515 (2004); D.~J.~Schwarz {\it et al.},
  \prl {\bf 93}, 221301 (2004).

\bibitem{gunder05} J.~Gundermann, astro-ph/0503014.

\bibitem{caillerie05} S.~Caillerie, M.~Lachi\`eze-Rey, A.~Riazuelo,
J.-P.~Luminet \& R.~Lehoucq, in preparation.

\bibitem{Rees68} M.J. Rees, Astrophys.\ J.\ {\bf 153}, L1 (1968).

\bibitem{HuWhite97} W.~Hu \& M.~White, New Astronomy, {\bf 2}, 323
(1997).

\bibitem{wmaphttp} WMAP website: http://map.gsfc.nasa.gov/~.

\bibitem{planckhttp} Planck website:
http://sci.esa.int/science-e/www/area/index.cfm?fareaid=17~.

\bibitem{bardeen} See for example V.~F.~Mukhanov, H.~A.~Feldman \&
R.~H.~Brandenberger, Phys.\ Rept.\ {\bf 215}, 203 (1992); R.~Durrer,
Fund.\ Cos.\ Phys.\ {\bf 14}, 209 (1994).

\bibitem{HuCMB} W.~Hu \& M.~White, \prd {\bf 56} 596 (1997).

\bibitem{revtopo} M.~Lachi\`eze-Rey \& J.-P.~Luminet, Phys.\ Rept.\
{\bf 254} 135 (1995).

\bibitem{lelalu96} R. Lehoucq, M. Lachi\`eze-Rey and J.-P. Luminet, 
Astron. Astrophys.  {\bf 313} 339 (1996).

\bibitem{detpola} J.~Kovac {\it et al.}, Nature (London) {\bf 420},
772 (2002).

\bibitem{reswmap} C.~L.~Bennett {\it et al.}, Astrophys.\ J.\ Suppl.\
{\bf 1}, 148 (2003).

\bibitem{aurich2005} R. Aurich, S. Lustig and F. Steiner, M.N.R.A.S. 
under press, astro-ph/0510847.

\end{thebibliography}
\end{document}